\documentstyle[prd,aps,epsfig,twocolumn]{revtex} % <-- FOR AUTHOR'S USE ONLY
\begin{document}
%\preprint{                                                  IASSNS-AST 96-31}
\draft
%............................................................................ 
\twocolumn[\hsize\textwidth\columnwidth%       <---    FOR AUTHOR'S USE ONLY
\hsize\csname @twocolumnfalse\endcsname%       <---
%............................................................................
\rightline{ IASSNS-AST 96/31, April 1996}
\rightline{\ \\[-2mm]}
\title{        Tests of three-flavor mixing in long-baseline
                    neutrino oscillation experiments}
\author{                     G.~L.~Fogli}
\address{     Dipartimento di Fisica dell'Universit{\`a}
             and Sezione INFN di Bari, I-70126  Bari, Italy}
\author{                       E.~Lisi}
\address{Institute for Advanced Study, Princeton, New Jersey 08540, U.S.A.\\
            and Dipartimento di Fisica dell'Universit{\`a}
             and Sezione INFN di Bari, I-70126 Bari, Italy\\[2mm] }
\maketitle
\begin{abstract}
%.............................................................................
We compare the effectiveness of various tests of three-flavor mixing in 
future long-baseline neutrino oscillation experiments. We analyze a 
representative case of mixing in a simplified three-flavor scheme, whose 
relevant parameters are one neutrino mass-square difference, $m^2$, and 
two mixing angles, $\psi$ and $\phi$. We show that an unambiguous 
determination of $\psi$ and $\phi$ requires flavor-appearance tests in 
accelerator experiments, as well as supplementary information from reactor 
experiments.
%.............................................................................
\end{abstract}
\vspace*{-0.2pc}
\vskip.9pc] %                      <---               FOR AUTHOR'S USE ONLY  
%-----------------------------------------------------------------------------
%-----------------------------------------------------------------------------
\newcommand{\InsertFigure}[1]{%
\epsfig{bbllx=2.5truecm,bblly=2.5truecm,bburx=20.5truecm,bbury=27.5truecm,%
height=13truecm,figure=#1}\vspace*{-27mm}}
%-----------------------------------------------------------------------------

%%%%%%%%%%%%%%%%%%%%%%%%%%%%%%%%%%%%%%%%%%%%%%%%%%%%%%%%%%%%%%%%%%%%%%%%%%%%%%
\section{Introduction}
%%%%%%%%%%%%%%%%%%%%%%%%%%%%%%%%%%%%%%%%%%%%%%%%%%%%%%%%%%%%%%%%%%%%%%%%%%%%%%

The next generation of proposed accelerator neutrino oscillation experiments
will be characterized by unprecedented source-detector distances:
730 km for the Fermilab to Soudan experiment (MINOS) \cite{MI95,MIok},
 250 km for the KEK-PS to SuperKamiokande experiment \cite{SKam},
68 km for the Brookhaven National Laboratory experiment E889, \cite{E889}
and 732 km for the CERN to Gran Sasso experiment, with ICARUS \cite{IC95},
RICH \cite{RICH}, or NOE \cite{NOE} as candidate detectors. Two reactor 
experiments under construction, Chooz \cite{Choo} and San Onofre 
\cite{SanO}, also have a much longer baseline (1 km) than  previous 
experiments of their class.

These long-baseline experiments are sensitive, in the favorable case of 
large $\nu$ mixing, to neutrino square mass differences as low as 
$10^{-3}$ eV$^2$. Moreover, they probe several flavor-oscillation channels. 
Oscillation signals in two or more independent channels could indicate flavor
mixing among three generations ($3\nu$ mixing).

In this report, we analyze the effectiveness of long-baseline experiments
in detecting $3\nu$ mixing signals. In Sec.~II we  describe a  representative 
case of mixing in a simple $3\nu$ scheme. (This scheme is extensively 
discussed in \cite{Fo95,Fo94}, to which we refer the reader for further 
details.)  In Secs.~III and IV we focus on oscillation searches at ICARUS 
and MINOS respectively. We also comment briefly upon searches at BNL~E889, 
Superkamiokande, RICH, and NOE. We show that flavor-appearance tests
are essential to constrain multiple interpretations of the signals.
In Sec.~V we show how reactor experiments may eliminate an ambiguity in
the determination of the mixing angles. We summarize our work in Sec.~VI.

%%%%%%%%%%%%%%%%%%%%%%%%%%%%%%%%%%%%%%%%%%%%%%%%%%%%%%%%%%%%%%%%%%%%%%%%%%%%%%
\section{Three-flavor mixing: A test case}
%%%%%%%%%%%%%%%%%%%%%%%%%%%%%%%%%%%%%%%%%%%%%%%%%%%%%%%%%%%%%%%%%%%%%%%%%%%%%%

In a three-generation approach, the $\nu$ states with definite flavor
$(\nu_\alpha=\nu_e,\,\nu_\mu,\,\nu_\tau)$ 
are a superposition of $\nu$ states with definite mass   
$(\nu_i=\nu_1,\,\nu_2,\,\nu_3)$: $\nu_\alpha = U_{\alpha i}\nu_i$,
$U$ being a unitary matrix. We assume \cite{Fo95} that the two 
independent neutrino square mass differences, 
$\delta m^2=m^2_2-m^2_1$ and $m^2=m^2_3-m^2_1$, 
are widely separated:
%................................................................ equation (1)
\begin{equation}
\delta m^2 \lesssim 10^{-4} {\rm\ eV}^2,\;
m^2 \gtrsim 10^{-3} {\rm\ eV}^2\;\longrightarrow\; 
\delta m^2 \ll m^2\ .
\end{equation}
%.............................................................................

Accelerator and reactor neutrino oscillations are then  simply described 
in terms of  $(m^2,\, U_{e3},\, U_{\mu3},\, U_{\tau3})$ or, equivalently, 
in terms of $(m^2,\,\psi,\,\phi)$, along with the parameterization 
($s_\phi=\sin\phi$, etc.): 
%................................................................ equation (2)
\begin{equation}
U^2_{e3}=s^2_\phi,\qquad U^2_{\mu3}=c^2_\phi s^2_\psi,\qquad 
U^2_{\tau3}=c^2_\phi c^2_\psi\ .
\end{equation}
%.............................................................................

The oscillation probabilities are given by \cite{Fo95}:
%................................................................ equation (3)
\begin{equation}
   P(\nu_\alpha\to\nu_\beta) =
\left\{\begin{array}{ll}
   4U^2_{\alpha 3}U^2_{\beta 3}S          &    {\rm\ if\ } \alpha\neq\beta
\\[1.5mm]
   1-4U^2_{\alpha 3}(1-U^2_{\alpha 3})S   &     {\rm\ if\ } \alpha=\beta
\end{array}\right.\ ,
\end{equation}
%.............................................................................
where $S=\sin^2(1.27 m^2 L/E_\nu)$, with $m^2$, $L$, and $E_\nu$
measured in eV$^2$, km, and GeV respectively. In this scheme, 
$\nu$ and $\bar\nu$ oscillations in vacuum are indistinguishable.

It has been shown  \cite{Fo95} that a  useful representation of the 
parameter space $(m^2,\,\psi,\,\phi)$ is obtained through
 $(\tan^2\psi,\,\tan^2\phi)$ sections at fixed values of $m^2$. A generic 
point in the $(\tan^2\psi,\,\tan^2\phi)$ plane represents the state 
$\nu_3$; the lower left and right corners and the upper side represent, 
asymptotically, $\nu_\tau$, $\nu_\mu$, and $\nu_e$ respectively.

For definiteness, we choose a representative $3\nu$ case:
%................................................................ equation (4)
\begin{equation}
\left\{\begin{array}{lcl}
         m^2&=&2.5\times10^{-2}{\rm\ eV}^2
\\[1mm]
        \nu_3&=&\sqrt{0.75}\,\nu_\tau+\sqrt{0.20}\,\nu_\mu+\sqrt{0.05}\,\nu_e
\end{array}\right.\ ,
\end{equation}
%............................................................................
corresponding to $\psi=27.3^{\circ}$ and  $\phi=12.9^{\circ}$. In Fig.~1,
this test case  is marked by a black circle $(\bullet)$ in the 
$(\tan^2\psi,\,\tan^2\phi)$ plane at $m^2=2.5\times10^{-2}$ eV$^2$.

The case in Eq.~(4) has the following  properties (see also \cite{Fo95}):
(1) it is allowed by the available reactor data, as shown in Fig.~1;
(2) it is in the sensitivity region of long-baseline experiments; 
(3) it is outside the sensitivity region of the running oscillation 
    experiments, CHORUS, NOMAD, KARMEN, and LSND (see \cite{Fo95} and 
    references therein);
(4) it is compatible with the indications of an anomalous $\mu/e$ flavor
    ratio in the atmospheric $\nu$ flux, and in particular with a
    value $R_{\mu/e}=(\mu/e)_{\rm data}/(\mu/e)_{\rm theory}\simeq 0.72$,
    as shown in Fig.~1; and
(5) the matrix elements $U_{\alpha3}$ are hierarchically ordered.
We will emphasize the results that remain valid in more general situations.

The adopted value of  $m^2$ is sufficiently high to allow in Eq.~(3), for 
our purposes, the ``averaged oscillation'' approximation,  
$\langle S\rangle \simeq1/2$. In a more refined approach,  $S$ has to
be computed using  the $E_\nu/L$  spectrum. Spectral analyses are also
necessary to infer the value of $m^2$. A discussion of such tests is beyond
the scope of this work.

Our strategy is the following.  For any experiment we define one or more 
observables, $Q_i$, that differ from zero in the presence of  oscillations. 
We assume a conservative, plausible estimate of $\pm 20\%$ for the total 
experimental errors, $\delta Q_i$, of the measured  signals $Q_i$. 
The actual values of the errors are not decisive for our considerations. 
The  $\delta Q_i$'s  will  be precisely estimated when the experiments will 
perform real measurements. 
In synthesis:
%................................................................ equation (5)
\begin{equation}
\left\{ \begin{array}{lcl}
Q_i   =  0 & \to & {\rm no\ oscillation} \\
Q_i \neq 0 & \to & {\rm oscillation\ \ }(\delta Q_i/Q_i\simeq\pm20\%)
\end{array}\right.\ .
\end{equation}
%.............................................................................

We then compute the  $Q_i$'s in the representative $3\nu$ test case 
[Eq.~(4)] using the averaged oscillation probabilities,
$P_{\alpha\beta}=\langle P(\nu_\alpha\to\nu_\beta)\rangle$ 
[Eq.~(3)]. Our goal is to study  how  the starting test values of the mixing 
angles may be inferred from measurements of the obervables $Q_i$.
 
%%%%%%%%%%%%%%%%%%%%%%%%%% INSERTION OF FIGS 1 AND 2 HERE (AUTHOR'S USE ONLY)
%%%%%%%%%%%%%%%%%%%%%%%%%%%%%%%%%%%%%%%%%%%%%%%%%%%%%%%%%%%%%%%%%%%%%%%%%%%%%%
\vspace*{-14mm}
\InsertFigure{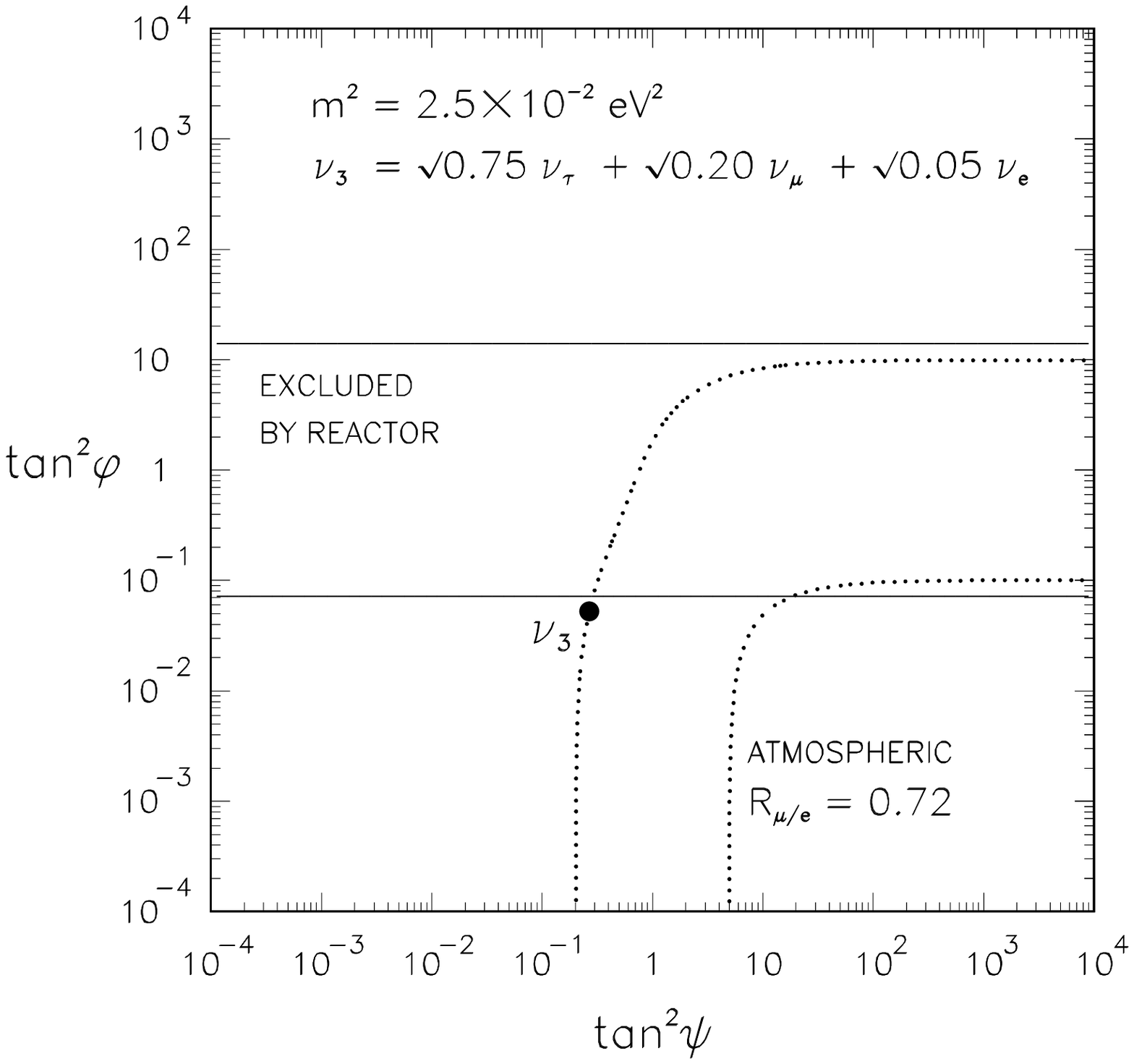}
\begin{figure}
\caption{The representative three-flavor mixing case ($\bullet$) in the 
	$(\tan^2\psi,\,\tan^2\phi)$ plane. The horizontal band is excluded 
	by present reactor $\nu$ data. The  $\Gamma$-shaped  region is 
	consistent with the indications of the atmospheric neutrino flavor 
	anomaly ($R_{\mu/e}=0.72$ along the dotted lines). See also 
	Ref.~\protect\cite{Fo95}.}
\end{figure}
%%%%%%%%%%%%%%%%%%%%%%%%%%%%%%%%%%%%%%%%%%%%%%%%%%%%%%%%%%%%%%%%%%%%%%%%%%%%%
%%%%%%%%%%%%%%%%%%%%%%%%%%%%%%%%%%%%%%%%%%%%%%%%%%%%%%%%%%%%%%%%%%%%%%%%%%%%%
\vspace*{-20mm}
\InsertFigure{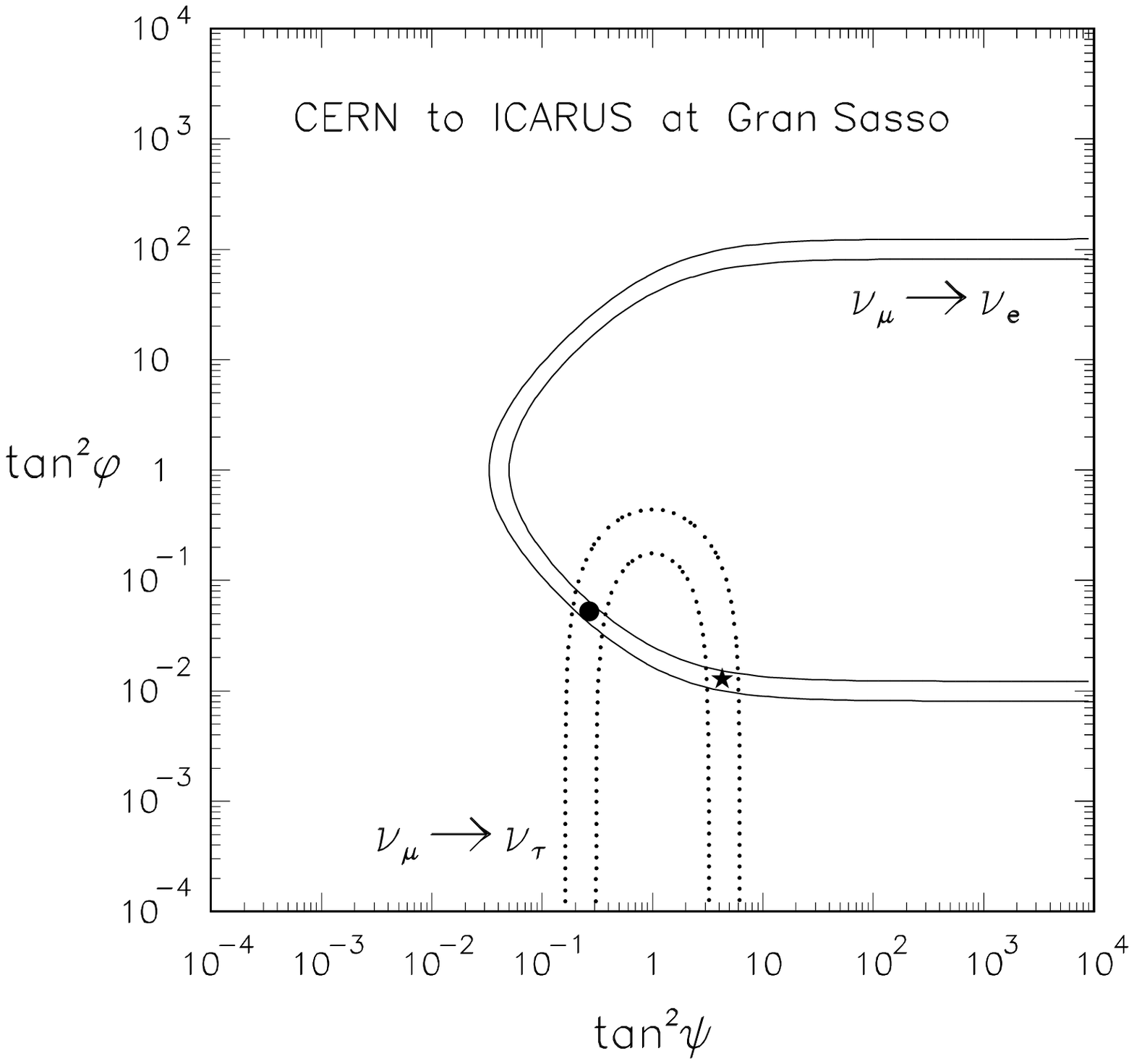}
\begin{figure}
\caption{Allowed bands corresponding to $\nu_\mu\to\nu_e$ and 
	$\nu_\mu\to\nu_\tau$ appearance searches at the ICARUS  experiment, 
	in the test case $(\bullet)$ of Fig.~1. The two bands also intersect 
	at a second point $(\star)$, giving rise to a twofold ambiguity in 
	the determination of the mixing angles $\psi$ and $\phi$.}
\end{figure}
%%%%%%%%%%%%%%%%%%%%%%%%%%%%%%%%%%%%%%%%%%%%%%%%%%%%%%%%%%%%%%%%%%%%%%%%%%%%%

%%%%%%%%%%%%%%%%%%%%%%%%%%%%%%%%%%%%%%%%%%%%%%%%%%%%%%%%%%%%%%%%%%%%%%%%%%%%%
\section{Oscillation tests at ICARUS}
%%%%%%%%%%%%%%%%%%%%%%%%%%%%%%%%%%%%%%%%%%%%%%%%%%%%%%%%%%%%%%%%%%%%%%%%%%%%%

The ICARUS detector can identify $e$-events downstream 
from a $\nu_\mu$ beam, 
thus probing $\nu_\mu\to\nu_e$  appearance. There are good 
prospects for direct $\tau$ identification. So we also assume that  
$\nu_\mu\to\nu_\tau$  appearance searches can be performed with the ICARUS 
detector.

Two observables can be defined in terms of the number of events
$N_{e,\mu,\tau}$ at the detector:
%............................................................. equations (6,7)
\begin{eqnarray}
Q_{\mu e}&\equiv&\frac{N_e({\rm observed})}{N_\mu({\rm expected})}
=P_{\mu e}\ , \\[1.5mm] 
Q_{\mu\tau}&\equiv&\frac{N_\tau({\rm observed})}{N_\mu({\rm expected})}
=P_{\mu\tau}\ .
\end{eqnarray}
%.............................................................................
In the $3\nu$ test case [Eq.~(4)], the above observables take the values
$Q_{\mu e}=0.02(1\pm0.2)$ and $Q_{\mu \tau}=0.30(1\pm0.2)$.

In Fig.~2 we show the corresponding curves of iso-$Q_{\mu e}$ and 
iso-$Q_{\mu \tau}$. The width of the bands is proportional to $\delta Q_i$. 
The bands intersect at two points. One represents, of course, the ``true'' 
test case $(\bullet)$. The second point $(\star)$ represents a ``false'' 
combination of mixing angles. In other words,  the determination of 
$(\psi,\,\phi)$ through the values $(Q_{\mu e},\,Q_{\mu \tau})$ is subject 
to a twofold ambiguity.

The appearance of a second  $3\nu$ solution is not limited to the test case, 
and cannot be avoided by reducing the errors $\delta Q_i$. In fact, the 
twofold solution arises essentially  from the quadratic mixing terms in 
Eq.~(3).

%%%%%%%%%%%%%%%%%%%%%%%%%%%%%%%%%%%%%%%%%%%%%%%%%%%%%%%%%%%%%%%%%%%%%%%%%%%%%%
\section{Oscillation tests at MINOS}
%%%%%%%%%%%%%%%%%%%%%%%%%%%%%%%%%%%%%%%%%%%%%%%%%%%%%%%%%%%%%%%%%%%%%%%%%%%%%%

The MINOS detector can identify $\mu$ tracks downstream from a  
$\nu_\mu$-beam. 
It is also  expected to identify electrons and  thus to be as sensitive 
to $Q_{\mu e}$ as the ICARUS detector. First we describe  two  oscillation 
search methods that use only $\mu$-identification, the $T$-test and  
the far/near-test \cite{MI95}. Then we discuss the $e$-appearance 
observable $Q_{\mu e}$.

The $T$ and  far/near tests  exploit the unoscillated measurements at a 
second, smaller  detector to be placed near  the $\nu$-beam source.
One separates the  events with ``one muon track'' ($1\mu$) and with 
``no muon 
track'' ($0\mu$). 

At the near detector, the number of events $n_{1\mu}$ and $n_{0\mu}$ 
are proportional to the $\nu_\mu$ charged (CC) and neutral current
(NC) $E_\nu$-averaged cross sections, $\sigma_{\rm C}$ and $\sigma_{\rm N}$:
%............................................................... equation (8)
\begin{equation}
n_{1\mu} \propto \sigma_{\rm C}\quad ,\quad
n_{0\mu} \propto \sigma_{\rm N}\quad.
\end{equation}
%............................................................................

At the far detector one also has to include the interactions of $\nu_e$'s 
and $\nu_\tau$'s. Downstream, $1\mu$ events are produced by:
(1) $\nu_\mu$ CC interactions, and 
(2) $\nu_\tau$ CC interactions followed by $\tau\to\mu$ decay. 
The $0\mu$ events are produced by: 
(1) $\nu_\mu$ NC interactions; 
(2) $\nu_e$ CC or NC interactions;
(3) $\nu_\tau$ NC interactions; and 
(4) $\nu_\tau$ CC interactions followed by $\tau\rightarrow x$ decay, 
    $x\neq\mu$.
Thus the number of events $N_{1\mu}$ and $N_{0\mu}$ in the far detector are 
%........................................................... equations (9a,9b)
\begin{mathletters}
\begin{eqnarray}
N_{1\mu}&\propto&
P_{\mu\mu}\, \sigma_{\rm C}+P_{\mu\tau}\,\eta\, \sigma_{\rm C}\, B\ , \\
N_{0\mu}&\propto& 
P_{\mu\mu}\,\sigma_{\rm N}+
P_{\mu e}(\sigma_{\rm N}+\sigma_{\rm C})+ \nonumber\\
& &P_{\mu\tau}[\sigma_{\rm N}+\eta\,\sigma_{\rm C}(1-B)]\ ,
\end{eqnarray}
\end{mathletters}
%.............................................................................
where $B=0.18$ is 
the branching ratio $(\tau\to\mu)/(\tau\to{\rm all})$,
and the factor $\eta\simeq0.31$ \cite{MI95}  accounts for the average 
reduction of the $\nu_\tau$ CC  cross section due to the $\tau$-mass.

Given Eqs.~(8,9), one introduces \cite{MI95} two ratios,
$t$ and $T$, at the near and far detector respectively: 
%........................................................... equations (10,11)
\begin{eqnarray}
t&\equiv&\frac{n_{1\mu}}{n_{1\mu}+n_{0\mu}}=
\frac{\sigma_{\rm C}}{\sigma_{\rm C}+\sigma_{\rm N}}\ ,
\\
T&\equiv&\frac{N_{1\mu}}{N_{1\mu}+N_{0\mu}}=
\frac{t\,(P_{\mu\mu}+\eta\,B\,P_{\mu\tau})}{1-t\,P_{\mu\tau}(1-\eta)}\ .
\end{eqnarray}
%.............................................................................
In deriving the R.H.S.\ of Eq.~(11) we have used the  unitarity property 
$P_{\mu e}+P_{\mu\mu}+P_{\mu\tau}=1$ and the definition of\ $t$ in Eq.~(10). 
At MINOS energies it is $t\simeq 0.72$ \cite{MI95}.

In the $T$-test,  $T\neq t$ signals neutrino oscillations.
A useful observable can be defined as:
%............................................................... equation (12)
\begin{equation}
Q_{T}\equiv1-\frac{T}{t}=
1-\frac{P_{\mu\mu}+\eta\,B\,P_{\mu\tau}}{1-t\,P_{\mu\tau}(1-\eta)}\ .
\end{equation}
%.............................................................................

In the far/near-test, one counts only $1\mu$  events at the two
detector distances, $L_{\rm far}$ and $L_{\rm near}$. Neutrino oscillations
are signaled by a non-zero value of the quantity:
%............................................................... equation (13)
\begin{equation}
Q_{f/n}\equiv 1-\frac{N_{1\mu}}{n_{1\mu}}\,
\frac{L^2_{\rm far}}{L^2_{\rm near}}=1-P_{\mu\mu}-P_{\mu\tau}\,\eta\,B\ .
\end{equation}
%.............................................................................

In the $3\nu$ test case,  $Q_{T}=0.18(1\pm0.2)$ and $Q_{f/n}=0.30(1\pm0.2)$.
The corresponding iso-signal bands are shown in Fig.~3. These $\Gamma$-shaped
bands largely superpose in the lower part, producing multiple ``false'' mixing
cases ($\circ\cdots\circ$) that have, within errors, the same values of 
$Q_{T}$ and $Q_{f/n}$ as the ``true'' test case ($\bullet$). The multiple 
solutions (except the highest in $\phi$) are allowed by present reactor data, 
and are consistent with the atmospheric $\nu$ anomaly (cf.\ Figs.~1 and 3). 
The larger the errors ($\delta Q_i$), the more ambiguous is the  $3\nu$ 
interpretation of the combined  $Q_{T}$ and $Q_{f/n}$ signals.

The  $T$ and far/near tests are not very effective in discriminating
among 3$\nu$ 
cases, since a large fraction of their signal is induced by a disappearance 
probability,  $P_{\mu\mu}$ [Eqs.~(12,13)].  If limited to $Q_{T}$ and 
$Q_{f/n}$,  the $3\nu$-mixing searches at MINOS may not be much more 
constraining than a pure $\mu$-disappearance experiment as BNL E889
\cite{E889}  (that, incidentally, is not proceeding \cite{MIok}).

Therefore it is very important that the MINOS detector 
probes $e$-appearance directly.  As shown in Fig.~3, the addition
of the $Q_{\mu e}$-test [Eq.~(6)] reduces the multiple solutions
($\circ\cdots\circ$) to the same two cases ($\bullet$
and $\star$) as in ICARUS (Fig.~2). The comparison of Figs.~2 and 3
shows that, unfortunately, the ambiguity is not solved by combining the 
MINOS and ICARUS tests. 

If direct $\tau$-identification were possible at MINOS
($\nu_\mu\to\nu_\tau$ appearance), the oscillation searches would be 
corroborated, since the  two solutions  ($\bullet$ and $\star$) would sit 
at the intersection of 4 bands in the $(\tan^2\psi,\,\tan^2\phi)$ plane---an
overconstrained situation. However, the $3\nu$-mixing ambiguity would  
still remain unsolved.

%%%%%%%%%%%%%%%%%%%%%%%%%%%%%%% INSERTION OF FIG. 3 HERE (AUTHOR'S USE ONLY)
%%%%%%%%%%%%%%%%%%%%%%%%%%%%%%%%%%%%%%%%%%%%%%%%%%%%%%%%%%%%%%%%%%%%%%%%%%%%%
\vspace*{-10mm}
\InsertFigure{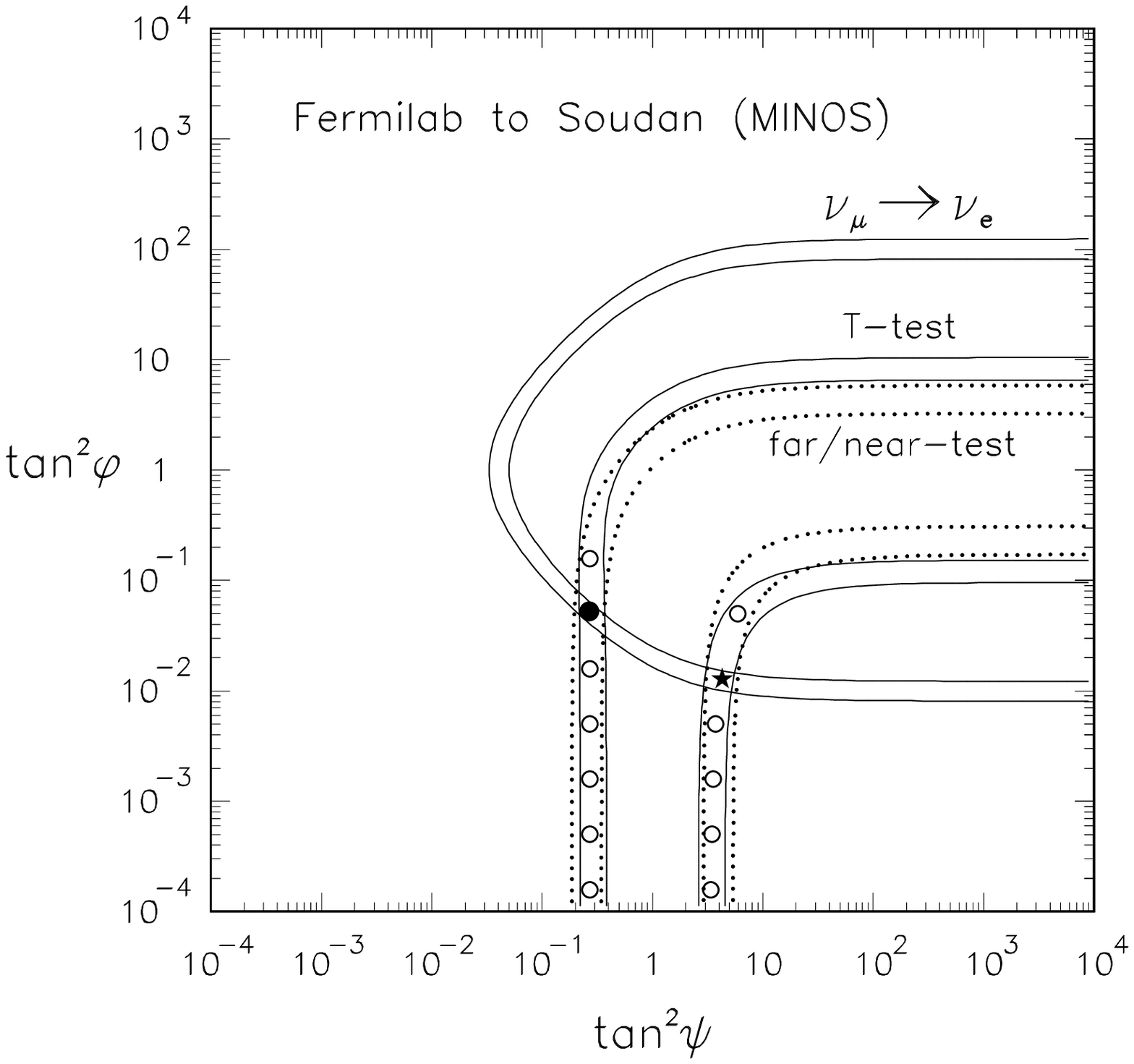}
\begin{figure}
\caption{Allowed bands corresponding to three oscillation tests
	at the MINOS experiments: $T$-test, far/near-test, and
	$\nu_\mu\to\nu_e$ appearance search. The combination 
	of the $T$ and far/near tests allows multiple solutions
	$(\circ\cdots\circ)$. Adding the $\nu_\mu\to\nu_e$ test, only 
	two solutions remain ($\bullet$ and $\star$) as in Fig.~2.}
\end{figure}
%%%%%%%%%%%%%%%%%%%%%%%%%%%%%%%%%%%%%%%%%%%%%%%%%%%%%%%%%%%%%%%%%%%%%%%%%%

The addition of atmospheric $\nu$ data in the usual  format of $\mu/e$ flavor 
ratio  is not particularly helpful, being very similar to the $T$-test (cf.\
Figs.~1 and 3). Flavor-separated  data \cite{FoLi} could be more effective 
in $3\nu$ mixing cases.

The two possible $\nu_\mu\to\nu_\mu$ and  $\nu_\mu\to\nu_e$ oscillation 
tests  at the planned SuperKamiokande experiment \cite{SKam} are 
similar to the MINOS  tests. However, the uncertainties \cite{SKam} are
expected to be larger, and the $3\nu$-mixing  discrimination power should 
be poorer.

As in MINOS, the RICH \cite{RICH} and NOE \cite{NOE} detectors
can distinguish $1\mu$ and $0\mu$ events and perform a
 $1\mu/(1\mu+0\mu)$ test. In  absence of a near-detector at CERN,
the normalization of their $\nu$-rates has to rely upon simulations.

%%%%%%%%%%%%%%%%%%%%%%%%%%%%%%%%%%%%%%%%%%%%%%%%%%%%%%%%%%%%%%%%%%%%%%%%%%%%%%
\section{Oscillation test at reactors}
%%%%%%%%%%%%%%%%%%%%%%%%%%%%%%%%%%%%%%%%%%%%%%%%%%%%%%%%%%%%%%%%%%%%%%%%%%%%%%

In reactor experiments, a single neutrino oscillation search 
is performed, namely the $\nu_e$-disappearance test:
%.............................................................. equation (14)
\begin{equation}
Q_{ee}\equiv 1-P_{ee}\ .
\end{equation}
%............................................................................

In the $3\nu$-mixing test case [Eq.~(4)], $Q_{ee}=0.095(1\pm0.2)$. This value 
is in the sensitivity region of the Chooz \cite{Choo} and San Onofre 
\cite{SanO} experiments.

In Fig.~4 we show the corresponding iso-signal stripes. It can be seen
how the supplementary reactor $\nu$ information singles out the ``true'' 
mixing case ($\bullet$) and solves the ICARUS and MINOS ambiguity. 

However, at slightly smaller test values of $\phi$  the reactor experiments 
become rapidly less sensitive and might not be able to select the  ``true'' 
solution within the errors.

%%%%%%%%%%%%%%%%%%%%%%%%%%%%%%%%%%%%%%% INSERT FIG 4 HERE (AUTHOR's USE ONLY)
%%%%%%%%%%%%%%%%%%%%%%%%%%%%%%%%%%%%%%%%%%%%%%%%%%%%%%%%%%%%%%%%%%%%%%%%%%%%%
\vspace*{-14.mm}
\InsertFigure{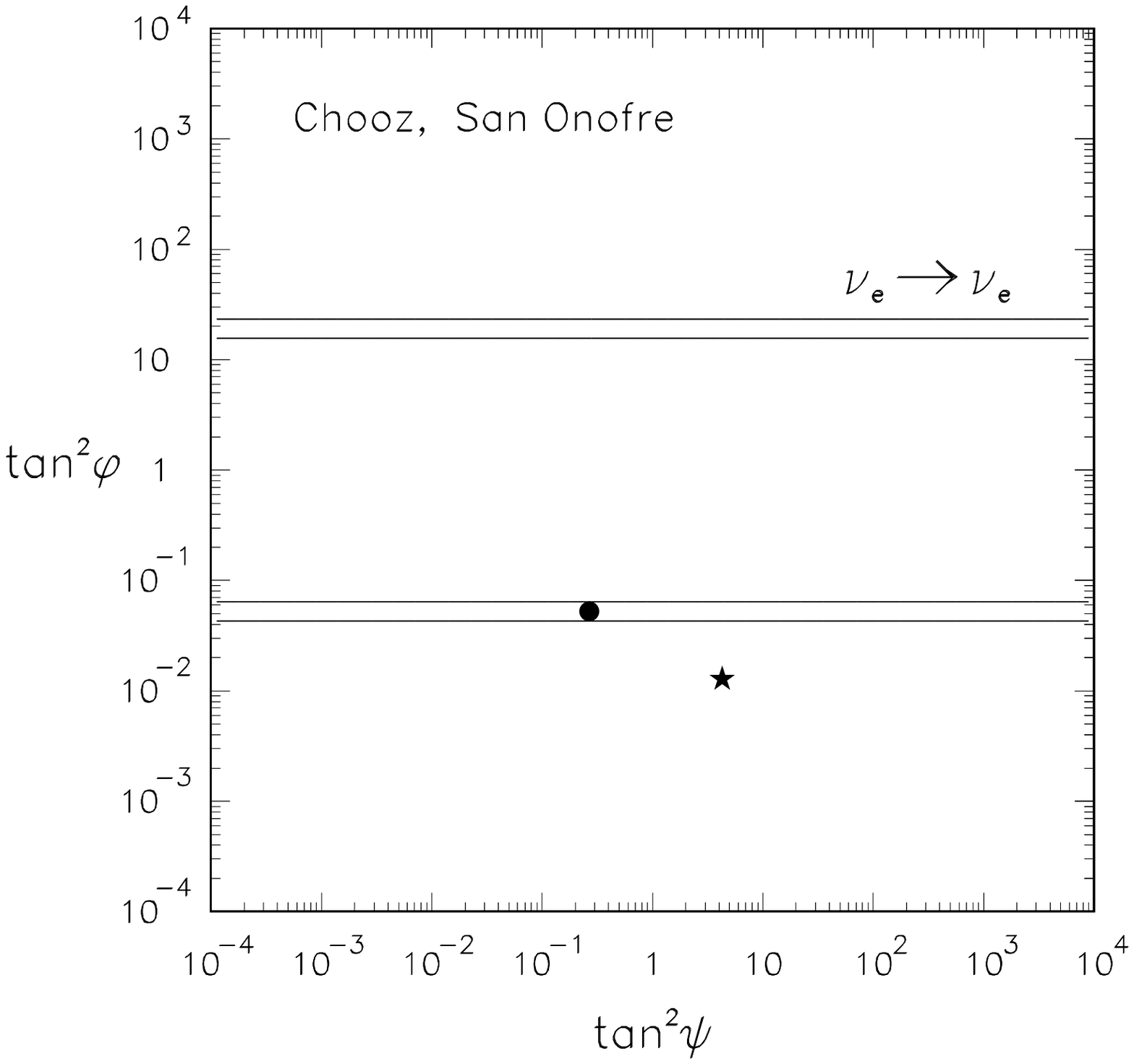}
\begin{figure}
\caption{Allowed bands corresponding to the $\nu_e\to\nu_e$ disappearance 
	test at reactors (Chooz, San Onofre). This test appears to solve 
	the twofold ambiguity between the ``true'' $(\bullet)$ and 
	``false'' $(\star)$ $3\nu$ mixing solution in  Figs.~2 and 3.}
\end{figure}
%%%%%%%%%%%%%%%%%%%%%%%%%%%%%%%%%%%%%%%%%%%%%%%%%%%%%%%%%%%%%%%%%%%%%%%%%%%%%

%%%%%%%%%%%%%%%%%%%%%%%%%%%%%%%%%%%%%%%%%%%%%%%%%%%%%%%%%%%%%%%%%%%%%%%%%%%%%
\section{Summary}
%%%%%%%%%%%%%%%%%%%%%%%%%%%%%%%%%%%%%%%%%%%%%%%%%%%%%%%%%%%%%%%%%%%%%%%%%%%%%

We have studied tests of three-flavor mixing in future long-baseline 
$\nu$ oscillation experiments. We have adopted a specific, simple 
$3\nu$-mixing case [Eq.~(4)], highlighting the aspects that have a more 
general validity.

We have considered five observables, 
$Q_{\mu e}$,
$Q_{\mu \tau}$,
$Q_{T}$,
$Q_{f/n}$, and
$Q_{ee}$,
that assume non-zero values in the presence of oscillations. 
A large fraction of the $Q_{T}$ and $Q_{f/n}$ oscillation signals (MINOS) 
is related to $\mu$-disappearance, and their combination  
may not be very effective in constraining $3\nu$-mixing cases.
The flavor-appearance observables $Q_{\mu \tau}$ (ICARUS) and $Q_{\mu e}$ 
(ICARUS, MINOS) allow a better resolution of $3\nu$ mixing angles, modulo a 
residual twofold ambiguity. The addition of the reactor observable 
$Q_{ee}$ (Chooz, San Onofre) may solve the ambiguity and select a unique 
$3\nu$  signal.

We thank  J.~H.~Cobb, M.~Goodman, P.~I.~Krastev, and D.~Petyt for 
reading the manuscript and
for useful information and suggestions. We thank the organizers of the Gran 
Sasso Long-baseline Workshop (Gran Sasso, Italy, 1995),
where preliminary results of this work were presented.
This work was supported  by INFN and by the Institute for Advanced Study.

%%%%%%%%%%%%%%%%%%%%%%%%%%%%%%%%%%%%%%%%%%%%%%%%%%%%%%%%%%%%%%%%%%%%%%%%%%%%%%%

\end{document}